\documentclass[]{spie}  

\usepackage{amsmath,amsfonts,amssymb}
\usepackage{graphicx}
\usepackage[colorlinks=true, allcolors=blue]{hyperref}

\title{The integration of the SOXS control electronics \\ towards the PAE}

\author[a]{M. Colapietro}
\author[a]{S. D'Orsi}
\author[a]{G. Capasso}
\author[a]{S. Savarese}
\author[a]{P. Schipani}
\author[a]{L. Marty}
\author[a]{R. Zanmar Sanchez}
\author[b]{M. Aliverti}
\author[c]{F. Battaini}
\author[c]{S. Di Filippo}
\author[c]{K. Radhakrishnan}
\author[c]{D. Ricci}
\author[c]{B. Salasnich}
\author[b]{S. Campana}
\author[c]{R. Claudi}
\author[d]{J. A. Araiza-Durán}
\author[c]{A. Baruffolo}
\author[e,f]{S. Ben-Ami}
\author[e]{A. Bichkovsky}
\author[d,g]{A. Brucalassi}
\author[h,i]{R. Cosentino}
\author[j]{F. D'Alessio}
\author[b]{P. D'Avanzo}
\author[i]{R. Di Benedetto}
\author[b]{M. Genoni}
\author[e]{O. Hershko}
\author[k,l]{H. Kuncarayakti}
\author[c]{L. Lessio}
\author[i]{E. Martinetti}
\author[i]{A. Miccichè}
\author[i]{G. Nicotra}
\author[m]{G. Pignata}
\author[n]{A. Rubin}
\author[o,i]{S. Scuderi}
\author[j]{F. Vitali}
\author[p]{J. Achrén}
\author[q]{I. Arcavi}
\author[b]{L. Asquini}
\author[e]{R. Bruch}
\author[c]{E. Cappellaro}
\author[a]{M. Della Valle}
\author[e]{A. Gal-Yam}
\author[h]{M. Hernandez Díaz}
\author[l,k]{J. Kotilainen}
\author[b]{M. Landoni}
\author[r]{G. Li Causi}
\author[k]{S. Mattila}
\author[i]{M. Munari}
\author[h]{H. Pérez Ventura}
\author[e]{M. Rappaport}
\author[b]{M. Riva}
\author[s,t]{S. Smartt}
\author[u]{M. Stritzinger}
\author[t]{D. Young}
\affil[a]{INAF - Osservartorio Astronomico di Capodimonte, Naples, Italy}
\affil[b]{INAF - Osservatorio Astronomico di Brera, Merate, Italy}
\affil[c]{INAF - Osservatorio Astronomico di Padova, Padua, Italy}
\affil[d]{INAF - Osservatorio Astronomico di Arcetri, Florence, Italy}
\affil[e]{Weizmann Institute of Science, Rehovot, Israel}
\affil[f]{Harvard-Smithsonian Center for Astrophysics, Cambridge, USA}
\affil[g]{Universidad Andres Bello, Santiago, Chile}
\affil[h]{INAF - Fundacion Galileo Galilei, Brena Baja, Spain}
\affil[i]{INAF - Osservatorio Astrofisico di Catania, Catania, Italy}
\affil[j]{INAF - Osservartorio Astronomico di Roma, Rome, Italy}
\affil[k]{Tuorla Observatory, Department of Physics and Astronomy, University of Turku, Finland}
\affil[l]{FINCA - Finnish Centre for Astronomy with ESO, Turku, Finland}
\affil[m]{Instituto de Alta Investigación, Universidad de Tarapaca, Arica, Chile}
\affil[n]{European Southern Observatory, Garching, Germany}
\affil[o]{INAF - Istituto di Astrofisica Spaziale e Fisica Cosmica, Milano, Italy}
\affil[p]{Incident Angle Oy, Turku, Finland}
\affil[q]{Tel Aviv University, Tel Aviv, Israel}
\affil[r]{INAF - Istituto di Astrofisica e Planetologia Spaziali, Rome, Italy}
\affil[s]{University of Oxford, Oxford, UK}
\affil[t]{Queen's University Belfast, School of Mathematics and Physics, Belfast, UK}
\affil[u]{Aarhus University, Aarhus, Denmark}

\authorinfo{Further author information:\\E-mail: mirko.colapietro@inaf.it}

\begin{document} 
\maketitle

\begin{abstract}
 SOXS (Son Of X-Shooter) is the new single object spectrograph for the ESO New Technology Telescope (NTT) at the La Silla Observatory, able to cover simultaneously both the UV-VIS and NIR bands (350-2000 nm).\\
 The instrument is currently in the integration and test phase, approaching the Preliminary Acceptance in Europe (PAE) before shipment to Chile for commissioning.\\
 After the assembly and preliminary test of the control electronics at INAF - Astronomical Observatory of Capodimonte (Napoli), the two main control cabinets of SOXS are now hosted in Padova, connected to the real hardware.\\
 This contribution describes the final electronic cabinets layout, the control strategy and the different integration phases, waiting for the Preliminary Acceptance in Europe and the installation of the instrument in Chile.
\end{abstract}


\section{INTRODUCTION}
\label{sec:intro}  
SOXS\cite{soxs-schipani} is the new spectroscopic facility for the ESO 3.58m New Technology Telescope (NTT) at the La Silla Observatory, which will become one of the premier transient follow-up instruments in the Southern hemisphere with a highly flexible schedule\cite{soxs-asquini} managed by the consortium.\\
The Assembly Integration and Verification phase of SOXS followed a modular approach, as the consortium is geographically spread. All the main subsystems have been internally aligned and tested in their respective institutes and then shipped to Padova, where the integration with the instrument control electronics and software, as well as the full system tests, are currently ongoing\cite{soxs-kalyan}.\\
In this contribution, the final electronic cabinets layout, the control strategy and the different integration phases are presented, waiting for the start of the Preliminary Acceptance in Europe and the installation of the instrument in Chile.

\section{INSTRUMENT OVERVIEW}

\begin{figure} [ht]
   \begin{center}
   \begin{tabular}{c} 
   \includegraphics[height=8cm]{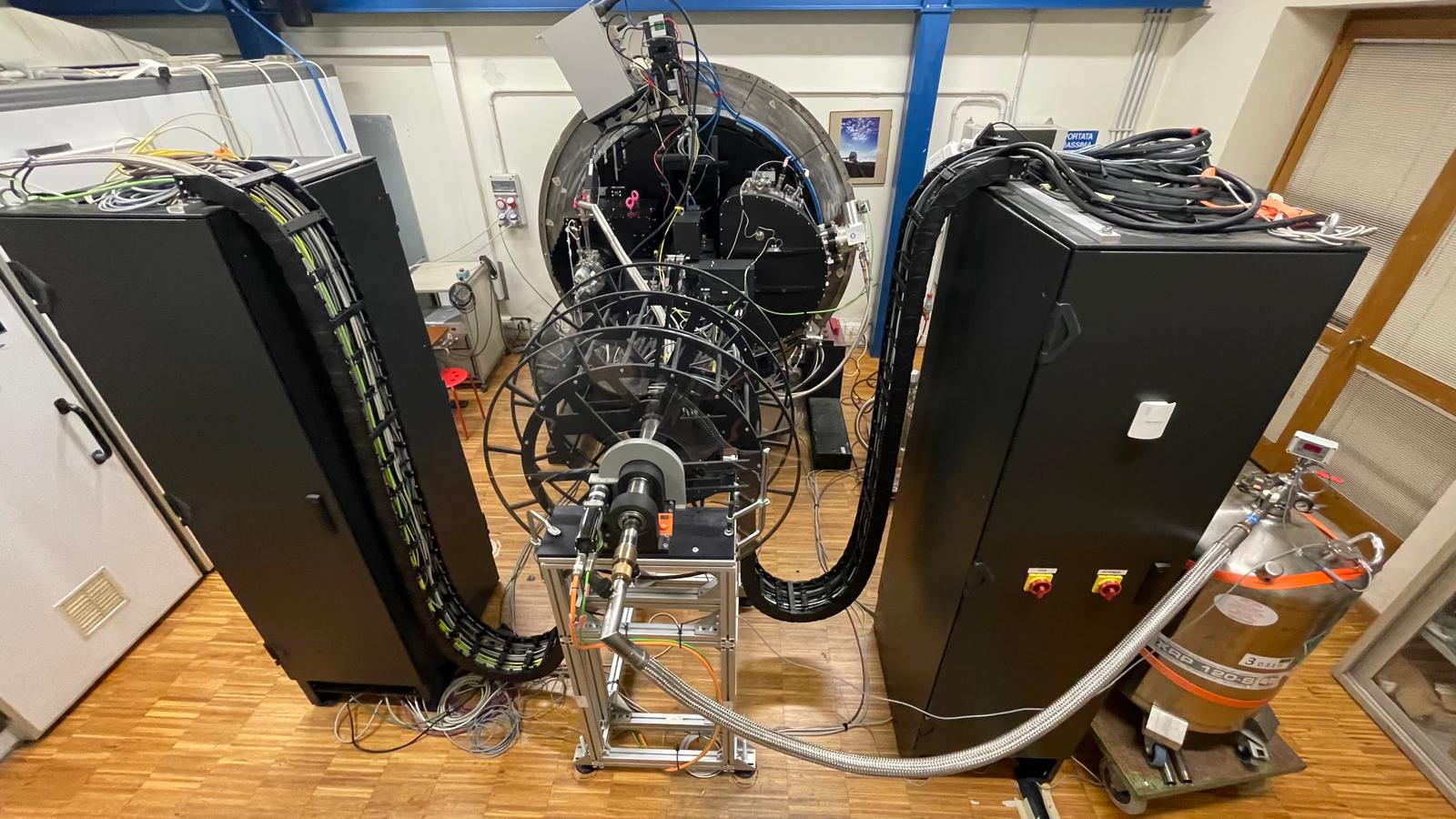}
   \end{tabular}
   \end{center}
   \caption { \label{fig:SOXS overview} 
SOXS instrument fully integrated in the Padua laboratory}
   \end{figure}

SOXS combines an average spectral resolution of ~4500 for a 1” slit with a wide spectral range (350-2000nm) obtained observing simultaneously the same target with two separate spectrographs, one operating in the UV-VIS (350-850nm) and the other one in the NIR (800-2000nm)\cite{soxs-genoni, soxs-vitali} wavelength regimes.\\
The instrument\cite{soxs-claudi} is completed by: a Common Path which represents the backbone of the system as it guides the light towards the two scientific arms; a Calibration Unit equipped with different light sources used for the daily calibrations of the two spectrographs; an Acquisition Camera\cite{soxs-araiza} intended for imaging and guiding functionalities.\\
All the five sub-systems are currently fixed on the SOXS flange and installed at the NTT simulator in the Padua integration laboratory (Figure \ref{fig:SOXS overview}).

\section{CONTROL ELECTRONICS ARCHITECTURE}
All the SOXS motorized functions, sensors and calibration lamps are controlled by the Instrument Control Electronics (ICE) and Software (ICS)\cite{soxs-ricci}.

The SOXS Control Electronics is mainly based on a PLC architecture. The communication between the PLC and the Instrument Workstation (IWS), that is used to manage the instrument, is obtained through the OPC-UA protocol in a server-client configuration: each device exposes, by means of the OPC-UA server installed on the PLC, a set of process variables that are accessible by the OPC-UA client available in the high-level control software running on the IWS.\\
An indipendent Cryo-Vacuum control System, based on a separate PLC, is responsible to manage and monitor all the valves, gauges and temperature sensors needed to guarantee the proper functionality of the two spectrographs.

All the controllers and devices are installed in specific subracks hosted in two Main Control Cabinets that are located at the two sides of the instrument platform and fed by the service connection points installed at the telescope.

\begin{figure} [ht]
   \begin{center}
   \begin{tabular}{c} 
   \includegraphics[height=10.5cm]{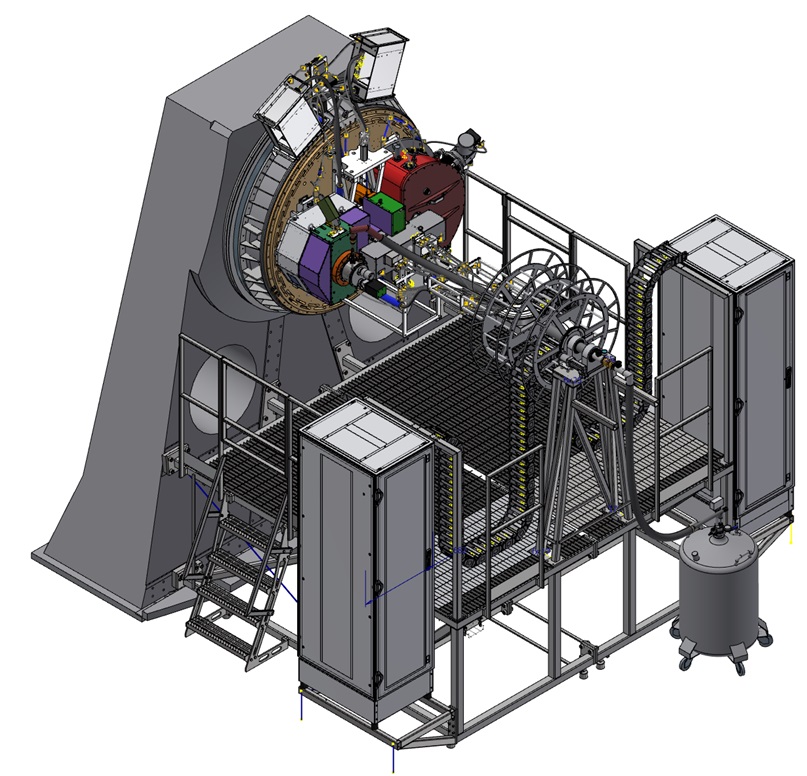}
   \end{tabular}
   \end{center}
   \caption { \label{fig:SOXS CAD} 
Final configuration of the SOXS instrument at the ESO NTT}
   \end{figure}

The two ESO NGC controllers foreseen to manage both the UV-VIS\cite{soxs-cosentino} and NIR detectors are installed directly on the flange due to limitations on the connection cables: their power supply units and the detector workstations are installed in the electronic cabinets. The SOXS final configuration at the NTT is shown in Figure \ref{fig:SOXS CAD}.

\section{ELECTRONICS INTEGRATION PHASES}
\label{sec:sections}
 
\subsection{Assembly phase}
\begin{figure} [ht]
   \begin{center}
   \begin{tabular}{c} 
   \includegraphics[height=12cm]{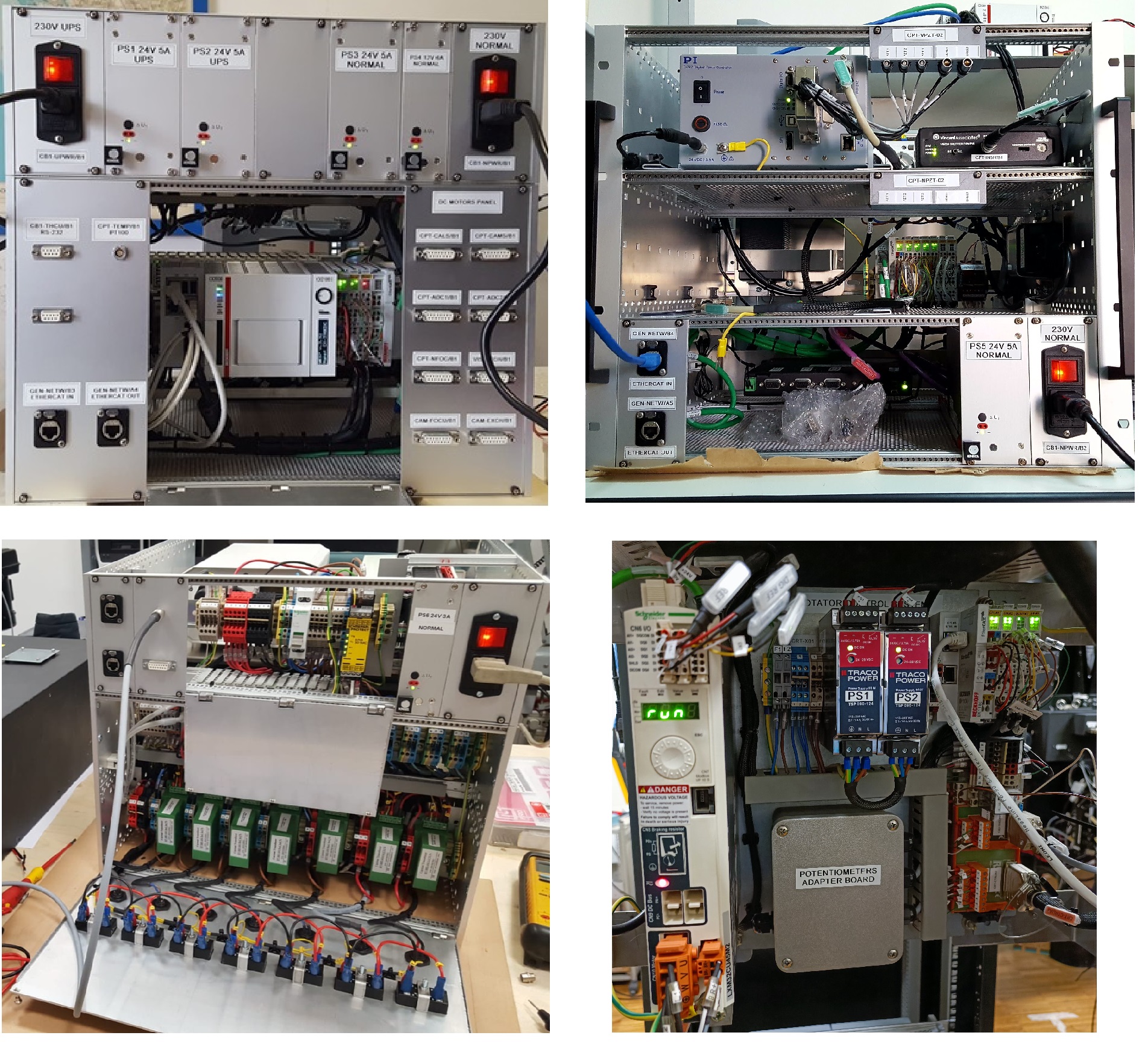}
   \end{tabular}
   \end{center}
   \caption { \label{fig:SOXS subracks} 
View of the Co-rotator control electronics system }
   \end{figure}

The SOXS Control Electronics\cite{soxs-capasso, soxs-colapietro} was mainly assembled and tested at the INAF Capodimonte observatory in Naples. It is characterized by four independent subracks (Figure \ref{fig:SOXS subracks}) containing all the PLC terminals to manage the instrument devices.

A first main subrack hosts the SOXS CPU and the PLC terminals to control the motorized functions of the different sub-systems: on the front side, a control touch-panel is foreseen to locally control the devices using dedicated GUIs during maintenance procedures; all the power, network and control interface connectors are installed on the back side. The connection between the electronics and the instrument is ensured by extension cables passing through the cable chains.

A second subrack has been design to be accessible only from the front side to optimize the volumes inside the cabinets. It hosts all the dedicated controllers needed to manage the instrument shutter and the piezo actuators, together with their power supply units. The instrument shutter controller interfaces with the PLC through digital I/O terminals. The piezo controllers, instead, need to be managed through a serial communication interfaces: custom software function blocks have been developed to properly control these actuators.

The Calibration Unit devices are managed by an additional subrack hosting all the lamp power supply units and the PLC I/O terminals to control the ON/OFF switching, monitor the status of the lamps and manage the motorized stage that allows to exchange between the two operating modes of the sub-system (full illumination and pinhole).

A last sub-unit manages the co-rotator control system. A dedicated drive adjusts the co-rotator motor speed according to the feedback signals coming from two linear potentiometers that provide the differential position between the instrument and the corotator. System enabling is managed and monitored by digital PLC I/O terminals.

\subsection{First integration}
The subracks described above were installed in the two SOXS Main Control Cabinets, properly cabled and shipped to the INAF – Astronomical Observatory of Padua for the first integration with the SOXS sub-systems.
All the actuators and sensors have been tested both through the PLC dedicated GUIs and the VLT SW installed in a separated machine. The PID parameters of the motors, with the effective loads, have been properly tuned in the TwinCAT PLC software to meet the requirements of the instrument.

\begin{figure} [ht]
   \begin{center}
   \begin{tabular}{c} 
   \includegraphics[height=6cm]{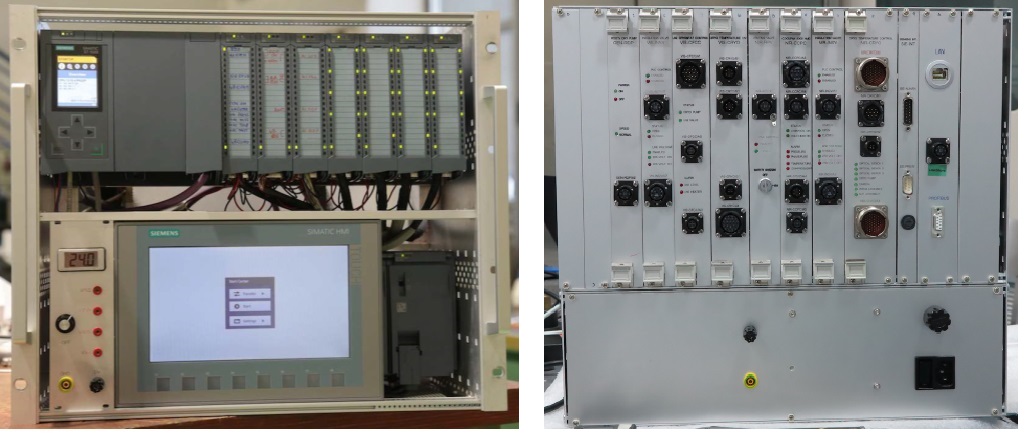}
   \end{tabular}
   \end{center}
   \caption { \label{fig:Cryo-vacuum electronics subrack} 
Front and rear view of the Cryo-vacuum control electronics subrack}
   \end{figure}

The Cryo-Vacuum control System (Figure \ref{fig:Cryo-vacuum electronics subrack}), assembled and tested independently in Catania, and the ESO NGC controllers together with their power supply units and workstations have been integrated in Padua afterwards. 

\begin{figure} [ht]
   \begin{center}
   \begin{tabular}{c} 
   \includegraphics[height=9cm]{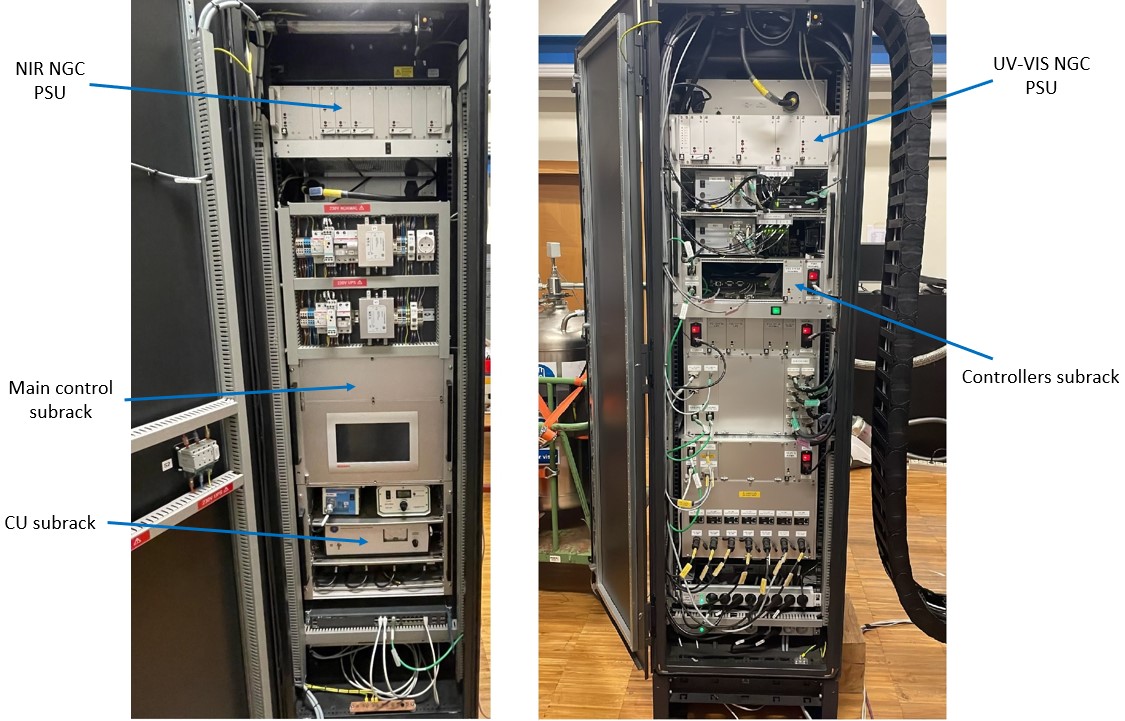}
   \end{tabular}
   \end{center}
   \caption { \label{fig:Cabinet1_views} 
First SOXS Main Control Cabinet: front (left) and rear (right) views}
   \end{figure}

\begin{figure} [ht]
   \begin{center}
   \begin{tabular}{c} 
   \includegraphics[height=10cm]{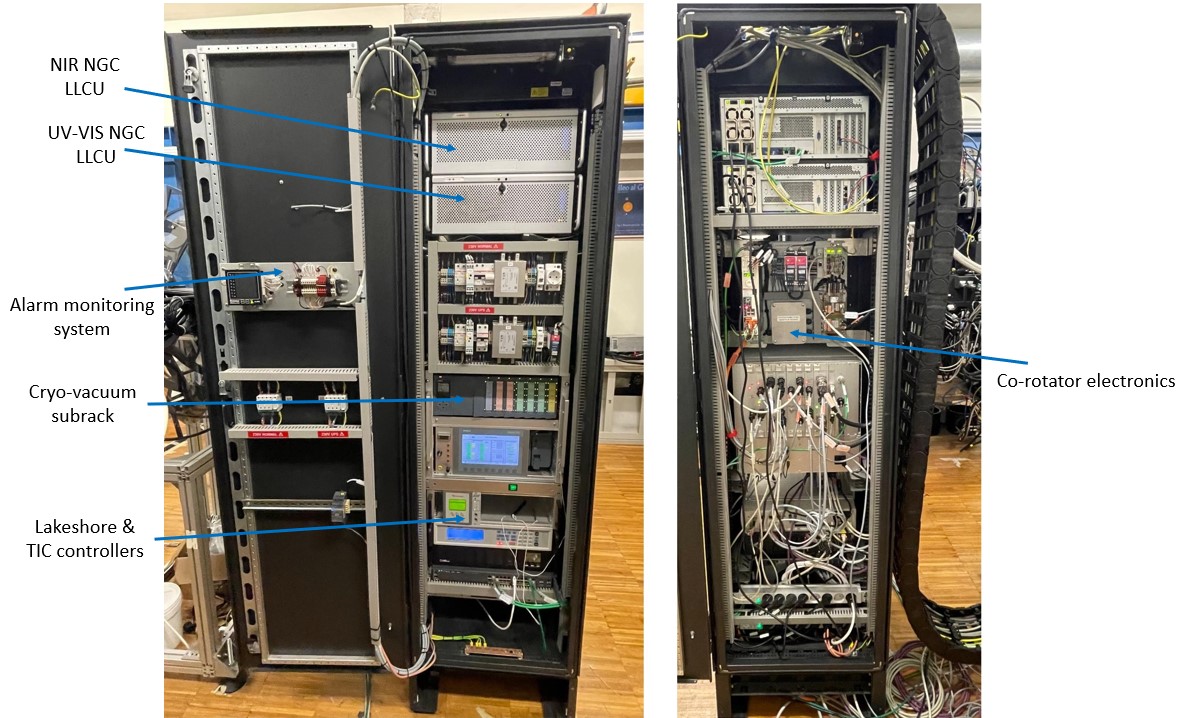}
   \end{tabular}
   \end{center}
   \caption { \label{fig:Cabinet2_views} 
Second SOXS Main Control Cabinet: front (left) and rear (right) views}
   \end{figure}

The SOXS electronic cabinets (Figure \ref{fig:Cabinet1_views} and Figure \ref{fig:Cabinet2_views}) are now in the final configuration and under tests, together with the Instrument Software, waiting for the Preliminary Acceptance in Europe process that will start next July.

\subsection{Final integration}
After the Preliminary Acceptance in Europe (PAE), the whole instrument will be dismounted and shipped to the ESO La Silla Observatory in Chile. The five subsystems will be checked and integrated again on the flange and fixed to the Nasmyth focus rotator of the NTT. The two electrical cabinets will be mounted at the two sides of the SOXS platform already installed, fed by the power and network services available at the telescope and all the cables connections will be restored.\\
All the functionalities of the instrument will be tested again during a second AIV phase, until SOXS become ready for the commissioning phase.

\section{CONCLUSIONS} 
During the assembly and integration phases, all the functionalities of the SOXS Control Electronics have been tested to verify that the design of the different sub-systems were correct.
The assembly of almost all the instrument control electronics took place at the INAF Capodimonte observatory (Naples).\\
Once every part was well mounted and assembled, the two Main Control Cabinets were sent to Padua and populated with the remaining parts of the SOXS Control Electronics assembled and tested independently in other institutes.\\
At this stage, the integration with all the instrument sub-systems was carried out: this allowed to test all the hardware interconnections and the PLC software proper operations.\\
All the motorized stages were mounted with the effective loads, and the PID parameters in the PLC software could be properly tuned in order to satisfy the requirements on positioning accuracy and repeatability.\\
The system is now in the final configuration and it is approaching to the PAE process: for this reason, it is subjected to intensive tests in order to verify the control system responses both in well working case and in a lot of possible failure cases.\\
Once the instrument will reach the NTT site in Chile, everything will be placed in its final position, all the cables connections will be restored and latest tests will be performed, waiting for the start of the instrument commissioning and first light.

\bibliography{main} 
\bibliographystyle{spiebib} 
\end{document}